# An Affine $SU(1,1)$ Description of Even-Even $^{106-122}Cd$ Isotopes in the Transitional Region of IBM

M. A. Jafarizadeh$^{a,b1}$, N. Fouladi$^c$, H. Sabri$^{c2}$, H. Fathi$^c$

$^a$Department of Theoretical Physics and Astrophysics, University of Tabriz, Tabriz 51664, Iran.

$^b$Research Institute for Fundamental Sciences, Tabriz 51664, Iran.

$^c$Department of Nuclear Physics, University of Tabriz, Tabriz 51664, Iran.

$^1$ E-mail: jafarizadeh@tabrizu.ac.ir
$^2$ E-mail: h-sabri@tabrizu.ac.ir

**Abstract**

In this paper, the properties of $^{106-122}Cd$ isotopes are considered in the $U(5) \leftrightarrow SO(6)$ transitional region of IBM1. With employing a transitional Hamiltonian which is based on affine $SU(1,1)$ Lie Algebraic technique, the energy levels and $B(E2)$ transition rates are calculated. The results are compared with the most recent experimental values where an acceptable degree of agreement is achieved. Also, the energy ratios, control parameters and empirical two neutron separation energies suggest spherical to $\gamma-$soft shape transitions in these nuclei where the $^{114}Cd$ and $^{122}Cd$ nuclei are considered as the best candidates for the $E(5)$ critical symmetry.

**Keywords:** Interacting Boson Model (IBM) – infinite dimensional algebra- energy levels- $B(E2)$ transition.

**PACS:** 21.60.Fw, 21.10.Re, 27.60.+j

**Introduction**

Atomic nuclei are known to exhibit changes of their energy levels and electromagnetic transition rates among them when the numbers of protons (or neutrons) are modified, resulting in the shape phase transitions [1-3] from one kind of collective behavior to another. In the framework of the Interacting Boson Model (IBM) [4-6], which describes nuclear structure of the even–even nuclei within the $U(6)$ symmetry, possessing the $U(5)$, $SU(3)$ and $O(6)$ dynamical symmetry limits, shape phase transitions have been studied 25 years ago using the classical limit of the model. These descriptions point out, there is a first order shape phase transition between $U(5)$ and $SU(3)$ limits and a second order shape phase transition between $U(5)$ and $O(6)$ limits. The analytic description of the structure at the critical point of the phase transition where an obvious change happened in the structure is under investigations. Numerical methods must be employed to diagonalize the Hamiltonian in these cases. Pan *et al* [7] proposed a new solution which based on affine $SU(1,1)$ algebraic technique which prepares the properties of nuclei classified in the $U(5) \leftrightarrow SO(6)$ transitional region of IBM [7-8].

It was long believed, the Cadmium isotopes were good examples of quadrupole vibrational nuclei, namely $U(5)$ nuclei [9-12]. However, during the last few years, new experimental data and calculations have led to a modified picture on these nuclei. These descriptions by particular emphasis on describing the experimental data via collective models suggest these nuclei to be soft with regard to the $\gamma$ deformation with an almost maximum effective trixiality of $\gamma \sim 30°$ [13]. On the other hand, the $^{110-122}Cd$ nuclei were studied by applying the IBM-1 in Ref. [14] where it was found that vibrations and intruder rotations coexist and the two collective modes are mixed in these nuclei. These mean, $Cd$ isotopes appear to evolve from $U(5)$ to $O(6)$-like structure in the IBM. Heyde *et al* in Ref.[16], performed an IBM-2 calculation for $^{110-114}Cd$ including the energy spectra and the $M1$ transitions. All these new experiments and theoretical calculations have provided new insights for these nuclei which is helpful to understand their structures [17-25]. Although, it is very difficult to treat them in terms of conventional mean field theories since they are neither vibrational nor rotational.

In this study, we investigate the even- even $^{106-122}Cd$ isotopes in the $U(5) \leftrightarrow SO(6)$ transition region and calculate the energy levels and $B(E2)$ transition probabilities in the frame work of IBM with using the affine $SU(1,1)$ algebraic technique. The estimated values for control parameters propose a spherical to $\gamma-$ soft shape transitions in these nuclei. The same shape transition is reveal by the evolution of two-neutron separation energies $S_{2n}$ [26] which investigated by experimental values [27-30]. Also, the $^{114}Cd$ and $^{122}Cd$ nuclei exhibits special values of control parameter and also energy ratios ($R_{4/2}$ and $R_{0_2^+/2_1^+}$) while suggest them as the best candidates for $E(5)$ dynamical symmetry.

This paper is organized as follows: section 2 briefly summarizes the theoretical aspects of transitional Hamiltonian and the affine $SU(1,1)$ algebraic technique, section 3 includes the numerical results obtained by applying the considered Hamiltonian to different isotopes. Section 4 is devoted to summarize and some conclusion based on the results given in section 3.

## 2. Theoretical framework

The phenomenological Interacting Boson Model (IBM) in terms of $U(5), O(6)$ and $SU(3)$ dynamical symmetries has been employed in describing the collective properties of several medium and heavy mass nuclei. These dynamical symmetries correspond to harmonic vibrator, axial rotor and $\gamma-$ unstable rotor as geometrical analogs respectively [4-6]. Although these symmetries are fairly successful in the investigation of low-lying nuclear states, the analytic description of the structure at the critical point of the phase transition is considered as issue, recently great analyses has been performed to describe them. Iachello [1-2] established a new set of dynamical symmetries for nuclei located at the critical points of transitional regions. The $E(5)$ symmetry which describe a second order phase transition correspond to the transitional states in the region from the $U(5)$ to the $O(6)$ symmetries in the IBM1. Different analyses [31-33] suggested some nuclei such as $^{134}Ba$, $^{104}Ru$ and etc as evidences for the empirical examples of such a symmetry. Some complicated numerical techniques are required to diagonalize the Hamiltonian in these transitional regions and critical points. An algebraic solution have been proposed by Pan *et al* [7-8] which based on the affine $SU(1,1)$ Lie algebra to exhibit the properties of nuclei located in the $U(5) \leftrightarrow SO(6)$ transitional region. Although the results of this approach are some what different from those of the IBM, but as have presented in Ref.[7], a clear correspondence with the description of the geometrical model for this transitional region is achieved.

## 2.1) The affine su(1,1) approach to the transitional Hamiltonian

The $SU(1,1)$ Algebra has been described in detail in Refs [7-8]. Here, we briefly outline the basic ansatz and summarize the results. The Lie algebra corresponding to the group $SU(1,1)$ is spanned by the three operators $\{S_1, S_2, S_0\}$,

$$[S_1, S_2] = -iS_0 \quad , \quad [S_2, S_0] = iS_1 \quad , \quad [S_0, S_1] = iS_2 \tag{2.1}$$

Representations of $SU(1,1)$ are determined by a single number $\kappa$, thus the representation of Hilbert space is spanned by orthonormal basis $|\kappa\mu\rangle$ where $\kappa$ can be any positive number and $\mu = \kappa, \kappa+1, \ldots$. Now, the infinite dimensional algebra is generated by using of [7-8]

$$S_n^{\pm} = c_s^{2n+1} S^{\pm}(s) + c_d^{2n+1} S^{\pm}(d) \quad , \quad S_n^0 = c_s^{2n} S^0(s) + c_d^{2n} S^0(d) \tag{2.2}$$

Where $c_s$ and $c_d$ are real parameters and $n$ can be $0, \pm 1, \pm 2, \ldots$. These generators satisfy the commutation relations

$$[S_m^0, S_n^{\pm}] = \pm S_{m+n}^{\pm} \quad , \quad [S_m^+, S_n^-] = -2S_{m+n+1}^0 \tag{2.3}$$

Then, the $\{S_m^{\mu}, \mu = 0, +, -; \pm 1, \pm 2, \ldots\}$ generates an affine Lie algebra $\widehat{SU(1,1)}$ without central extension. With employing the generators of $\widehat{SU(1,1)}$ Algebra, the following Hamiltonian for transitional region between $U(5) \leftrightarrow SO(6)$ limits is prepared;

$$\hat{H} = g\, S_0^+ S_0^- + \varepsilon\, S_1^0 + \gamma\, \hat{C}_2(SO(5)) + \delta\, \hat{C}_2(SO(3)) \tag{2.4}$$

$g, \varepsilon, \gamma$ and $\delta$ are real parameters. It can be shown that (2.4), would be equivalent with the SO(6) Hamiltonian if $c_s = c_d$, and with U(5) Hamiltonian when $c_s = 0$ & $c_d \neq 0$. Therefore, the $c_s \neq c_d \neq 0$ requirement just corresponds to the $U(5) \leftrightarrow SO(6)$ transitional region. In our calculation, we take $c_d$ (=1) constant value and $c_s$ vary between 0 and $c_d$.

For evaluating the eigenvalues of Hamiltonian (2.4), the eigenstates are considered as [7-8]

$$|k; v_s v n_\Delta LM\rangle = \sum_{n_i \in Z} a_{n_1} a_{n_2} \ldots a_{n_k} x_1^{n_1} x_2^{n_2} \ldots x_k^{n_k} S_{n_1}^+ S_{n_2}^+ \ldots S_{n_k}^+ |lw\rangle \quad , \tag{2.5}$$

Due to the analytical behavior of the wavefunctions, it suffices to consider $x_i$ near zero. With using the commutation relations between generators of $SU(1,1)$ Algebra (2.3), wavefunctions express as:

$$|k; v_s v n_\Delta LM\rangle = N S_{x_1}^+ S_{x_2}^+ \ldots S_{x_k}^+ |lw\rangle \quad , \tag{2.6}$$

where $N$ is the normalization factor and

$$S_{x_i}^+ = \frac{c_s}{1 - c_s^2 x_i} S^+(s) + \frac{c_d}{1 - c_d^2 x_i} S^+(d) \quad , \tag{2.7}$$

The c-numbers $x_i$ are determined by the following set of equations

$$\frac{\epsilon}{x_i} = \frac{gc_s^2(v_s + \frac{1}{2})}{1-c_s^2 x_i} + \frac{gc_d^2(v + \frac{5}{2})}{1-c_d^2 x_i} - \sum_{i \neq j} \frac{2}{x_i - x_j} \qquad \text{for i=1,2,...,k} \qquad (2.8)$$

The eigenvalues $E^{(k)}$ of Hamiltonian (2.4) can then be expressed;

$$h^{(k)} = \sum_{i=1}^{k} \frac{\varepsilon}{x_i}, \qquad (2.9)$$

Which

$$E^{(k)} = h^{(k)} + \gamma v(v+3) + \delta L(L+1) + \varepsilon \Lambda_1^0 \quad , \quad \Lambda_1^0 = \frac{1}{2}[c_s^2(v_s + \frac{1}{2}) + c_d^2(v_s + \frac{5}{2})] \qquad (2.10)$$

The quantum number ($k$) is related to the total boson number $N$ by

$$N = 2k + v_s + v$$

To obtain the numerical results for $E^{(k)}$ (energy spectra of considered nuclei), we have followed the prescriptions introduced in Refs.[7-8], namely, a set of non-linear Beth-Ansatz equations (BAE) with k-unknowns for k-pair excitations must be solved. Now let us, change variables as

$$\epsilon = \frac{\varepsilon}{g}(g = 1 \; kev \; [7\text{-}8]) \qquad c = \frac{c_s}{c_d} \leq 1 \qquad y_i = c_d^2 x_i$$

so, the new form of (2.8) would be

$$\frac{\epsilon}{y_i} = \frac{c^2(v_s + \frac{1}{2})}{1-c^2 y_i} + \frac{(v + \frac{5}{2})}{1-y_i} - \sum_{i \neq j} \frac{2}{y_i - y_j} \qquad \text{for i=1,2,...,k} \qquad (2.11)$$

To calculate the roots of Beth-Ansatz equations (BAE) with specified values of $v_s$ and $v$, we have solved equation (2.11) with definite values of $c$ and $\varepsilon$ for $i = 1$. Then, we used "Find root" in Maple13 to get all $y_j'$s. We carry out this procedure with different values of $c$ and $\varepsilon$ to provide energy spectra (after inserting $\gamma$ and $\delta$) with minimum variation in compare to experimental values;

$$\sigma = (\frac{1}{N_{tot}} \sum_{i, tot} |E_{exp}(i) - E_{cal}(i)|^2)^{1/2}$$

($N_{tot}$ is the number of energy levels where included in the fitting processes). The method for fixing the best set of parameters in the Hamiltonian ($\gamma$ and $\delta$) includes carrying out a least-square fit (LSF) of the excitation energies of selected states $0_1^+, 2_1^+, 4_1^+, 0_2^+, 2_2^+, 4_2^+$ and etc or the two neutron separation energies of considered nuclei with available experimental data [27-30].

## 2.2) $B(E2)$ Transition probabilities

The reduced electric quadrupole transition probabilities $B(E2)$ are considered as observable which as well as quadrupole moment ratios within the low-lying state bands prepare more information about the nuclear structure. The E2 transition operator must be a Hermitian tensor of rank two and consequently the number of bosons must be conserved. With these constraints, there are two operators possible in the lowest order, therefore, the electric quadrupole transition operator employed in this study is given by [7],

$$\hat{T}_\mu^{(E2)} = q_2 [\hat{d}^\dagger \times \tilde{s} + \hat{s}^\dagger \times \tilde{d}]_\mu^{(2)} + q_2' [\hat{d}^\dagger \times \tilde{d}]_\mu^{(2)} \quad , \qquad (2.12)$$

Where $q_2$ is the effective quadrupole charge and $q_2'$ is a dimensionless coefficient. Also, the reduced electric quadrupole transition rates between $I_i \rightarrow I_f$ states are given by

$$B(E2; I_i \to I_f) = \frac{\left|\langle I_f \| T(E2) \| I_i \rangle\right|}{2I_i + 1} \quad , \tag{2.13}$$

To consider the quantities $(q_2, q_2')$, we have employed the same method introduced in Ref.[7], namely, in the fitting procedures, these parameters would be described as a function of total boson number $(N)$.

## 3. Numerical result

### 3.1) Energy levels

The investigations of the experimental energy spectra [14-19], suggest the $^{106-122}Cd$ nuclei as the empirical evidences for the transitional region $U(5) \leftrightarrow SO(6)$. Consequently, the Hamiltonian of the transition region (2.4) has been employed. There are 12 levels up to the $2_4^+$ level for each nucleus included in the fitting procedure. The best fit for the parameters of Hamiltonian, namely $\varepsilon, c_s, \delta$ and $\gamma$ used in the present work are presented in Table 1 which evaluated by technique described in the previous section. These quantities describe the best agreement between the calculated energy levels in this approach and their corresponding experimental data taken from [27-30], it means, the minimum values for $\sigma$. Figure (1) describes the available experimental levels and their corresponding calculated levels for $^{106-108}Cd$ nuclei respectively in the low-lying region of spectra. An acceptable degree of agreement is apparent between them.

| Nucleus | $N$ | $\varepsilon(kev)$ | $c_s$ | $\gamma(kev)$ | $\delta(kev)$ | $\sigma$ |
|---|---|---|---|---|---|---|
| $^{106}_{48}Cd$ | 5 | 790 | 0.43 | −36.17 | 13.41 | 201 |
| $^{108}_{48}Cd$ | 6 | 750 | 0.41 | −31.02 | 11.07 | 148 |
| $^{110}_{48}Cd$ | 7 | 620 | 0.58 | −14.44 | 16.22 | 119 |
| $^{112}_{48}Cd$ | 8 | 650 | 0.57 | −21.21 | 19.58 | 92 |
| $^{114}_{48}Cd$ | 9 | 685 | 0.53 | −23.71 | 28.08 | 133 |
| $^{116}_{48}Cd$ | 8 | 900 | 0.37 | −47.94 | 2.18 | 125 |
| $^{118}_{48}Cd$ | 7 | 740 | 0.40 | −32.19 | 12.41 | 83 |
| $^{120}_{48}Cd$ | 6 | 710 | 0.42 | −25.06 | 17.21 | 104 |
| $^{122}_{48}Cd$ | 5 | 695 | 0.48 | −21.12 | 25.19 | 163 |

Table1.The parameters of Hamiltonian (2.4) prepared by LSF technique for different isotopes of $Cd$, $N$ describes the boson number and $\varepsilon, c_s, \gamma$ and $\delta$ are the parameters of transitional Hamiltonian (2.4) for each nuclei. $\sigma$ regards as the quality of fitting processes.

### 3.2) Transition probabilities

The stable even-even nuclei in the $Cd$ isotopic chain exhibit an excellent opportunity for studying the behavior of the total low-lying $E2$ strengths in the transitional region from deformed to spherical nuclei. Computation of electromagnetic transition is a sign of good test for the nuclear model wave functions. To determine the boson effective charges, we have employed the same method introduced in Ref.[7], namely, we perform a fit to the empirical $B(E2)$ values in such isotopes where in the fitting procedures, these parameters would be described as a function of total boson number(N). The matrix elements of the $E2$ operator of Eq.(2.12) have been calculated by using the following values of the effective charge parameters where presented in Table2.

| Nucleus | $q_2$ | $q_2^{'}$ | Nucleus | $q_2$ | $q_2^{'}$ |
|---|---|---|---|---|---|
| $^{106}_{48}Cd$ | 2.533 | −0.141 | $^{108}_{48}Cd$ | 2.555 | −0.209 |
| $^{110}_{48}Cd$ | 2.579 | −0.312 | $^{112}_{48}Cd$ | 2.602 | −0.466 |
| $^{114}_{48}Cd$ | 2.625 | −0.695 | $^{116}_{48}Cd$ | 2.594 | −0.501 |
| $^{118}_{48}Cd$ | 2.492 | −0.343 | $^{120}_{48}Cd$ | 2.398 | −0.229 |
| $^{122}_{48}Cd$ | 2.514 | −0.151 | | | |

Table2. The experimental values of $B(E2)$ for considered nuclei and the coefficients ($q_2, q_2^{'}$) employed in this analysis which estimated by similar technique have been described in Ref.[7].

A comparison between the results of the present analysis for $B(E2; 2_1^+ \to 0_1^+)$, $B(E2; 4_1^+ \to 2_1^+)$ and also the ratios of these quantities with experimental values [27-30] displayed in Figure 2, where a good agreements are apparent. In all figures of the present paper, the uncertainties of the experimental data which are smaller than the size of the symbols are not represented.

As it can be seen from theses figures and Tables, the calculated energy spectra in this approach are generally in good agreements with the experimental data. The calculated results shown in these figures indicate the elegance of the fits presented in this technique and they suggest the success of the guess in parameterization. Also, the calculated $B(E2)$ transition probabilities of some even-even $Cd$ isotopes by using the model perspectives, exhibit nice agreement with experimental ones.

These results give information on structural changes in the nuclear deformation and shape-phase transitions. The shape-phase transition was associated with a sudden change in the nuclear collective behavior reflected in the increase of $R_{4/2} = E_{4_1^+}/E_{2_1^+}$, i.e. rotational excitation. It means, this quantity varies from the spherical vibrator limit's value 2.0, to the deformed $\gamma$ − soft limit, 2.50 while Iachello in Refs.[1-2], proposed the 2.20 for $E(5)$ dynamical symmetry which describe the critical point of $U(5) \leftrightarrow SO(6)$ transitional region. Also, Iachello considered the vibrational excitation, i.e. $R_{0/2} = E_{0_2^+}/E_{2_1^+}$ to describe $E(5)$ critical symmetry while proposed 3.03 for the critical point of transitional region. On the other hand, the ground state two-neutron separation energies ($S_{2n}$) are very sensitive observables to the details of nuclear structure. Gross nuclear structure features, like major shell closures, are clearly seen in the evolution of this observable along the isotopic chains [34]. Zamfir *et al* [26] suggest a smooth variation of two neutron separation energies $(S_{2n})$ for nuclei which describe second order shape-phase transition between spherical –$U(5)$– and axial rotor –$SO(6)$– limits. The correlations of these observables, one $S_{2n}$ related to ground state properties and other, $R_{4_1^+/2_1^+}$ and $R_{0_2^+/2_1^+}$, related to the properties of the excited states, explore very clearly the shape-phase transition region. In order to highlight the nuclear structure information contained in this type of observables, we investigated the evolution of the experimental two-neutron separation energies ($S_{2n}$) along the isotopic chains for the even-even $Cd$ nuclei where presented in Figure3. We used for the present study the last review of the nuclear masses reported in Ref [26] and, where available, very recent data [27-30]. Figure3, confirm the predictions of Zamfir *et al* and suggest a second order phase transition for this chain of $Cd$ isotopes where estimated control parameters ($c_s$) suggest similar evolution between spherical ($c_s = 0$ or $U(5)$ limit) and $\gamma$–unstable($c_s = 1$ or $SO(6)$ limit) shapes as presented in Table3.

| Nuclei | $^{106}_{48}Cd$ | $^{108}_{48}Cd$ | $^{110}_{48}Cd$ | $^{112}_{48}Cd$ | $^{114}_{48}Cd$ | $^{116}_{48}Cd$ | $^{118}_{48}Cd$ | $^{120}_{48}Cd$ | $^{122}_{48}Cd$ |
|---|---|---|---|---|---|---|---|---|---|
| $c_s$ | 0.43 | 0.41 | 0.58 | 0.57 | 0.53 | 0.37 | 0.40 | 0.42 | 0.48 |
| $R_{4/2} = E_{4_1^+}/E_{2_1^+}$ Rotational excitation | 2.35 | 2.37 | 2.33 | 2.31 | 2.28 | 2.37 | 2.39 | 2.38 | 2.30 |
| $R_{0/2} = E_{0_2^+}/E_{2_1^+}$ vibrational excitation | 2.85 | 2.73 | 2.25 | 1.99 | 2.05 | 2.53 | 2.66 | 2.77 | 3.01 |

Table3. The values of control parameter $c_s$, $R_{4/2}$ and $R_{0/2}$ for considered nuclei where $^{114}Cd$ and $^{122}Cd$ describe closer approaches to the predictions of Iachello for $E(5)$ critical symmetry.

The values of control parameters $(c_s)$ propose the structural changes in the nuclear deformation and shape-phase transitions in even-even $^{106-124}Cd$, namely $c_s \sim 0 \rightarrow 1$. On the other hand, Iachello in Refs.[2] considered $\eta$ as control parameter in his description of shape-phase transition, while at and around $\eta \sim 0.5$, one can expect the critical points of transitional regions. It means, one can consider $E(5)$ symmetry by $c_s \sim 0.5$ in our approach. The properties of considered nuclei ($c_s$, $R_{4_1^+/2_1^+}$ and $R_{0_2^+/2_1^+}$) listed in Table3, suggest similar behavior where $^{114}Cd$ and $^{122}Cd$ isotopes which are known as the best candidates for $E(5)$ symmetry [13,19], exhibit special values of these quantities which are closer to the predictions of Iachello.

## 4. CONCLUSIONS

In this paper, with employing an affine $SU(1,1)$ lie algebra, the energies and $B(E2)$ transition probabilities of $^{106-122}Cd$ nuclei are described within the framework of the IBM. The validity of the presented parameters in the IBM formulations has been investigated and it is seen that there is an existence of a satisfactory agreement between the presented results and the experimental data. We may conclude that the general characteristics of the $Cd$ isotopes are well accounted in this study and the idea of shape coexistence in this region is supported. The elegance of Figures (1,2) suggest an acceptable agreement between the presented IBM-1 results, and experimental results for considered nuclei. The $Cd$ isotopes are close to both the proton and the neutron closed shell and these nuclei are not expected to be deformed. It can also be stated that gamma-soft rotor features exist in $Cd$ isotopes, but with a dominancy of vibrational character where $^{114}Cd$ and $^{122}Cd$ provide evidences for $E(5)$ symmetry. The obtained results in this study confirm that this technique is worth extending for investigating the nuclear structure of other nuclei existing around the mass of $A \sim 130$.

# Figure caption

**Figure1.** Comparison of the calculated energy levels and experimental spectra [27-30] for $^{106-108}Cd$. Due to the similar correspondences, we wouldn't present this comparison for other isotopes.

**Figure2.** Comparison of the calculated transition probabilities and corresponding experimental values [27-30] $B(E2; 2_1^+ \to 0_1^+)$ for $^{106-122}Cd$ isotopes, $B(E2; 4_1^+ \to 2_1^+)$ for $^{106-120}Cd$. Also, a comparison between the calculated $B_{4/2}$ ratios and experimental (available values) ones are presented for $^{106-120}Cd$ isotopes.

**Figure3.** The experimental $S_{2n}$ energies (in *kev*) for considered nuclei ($^{106-122}Cd$).

Figure1.

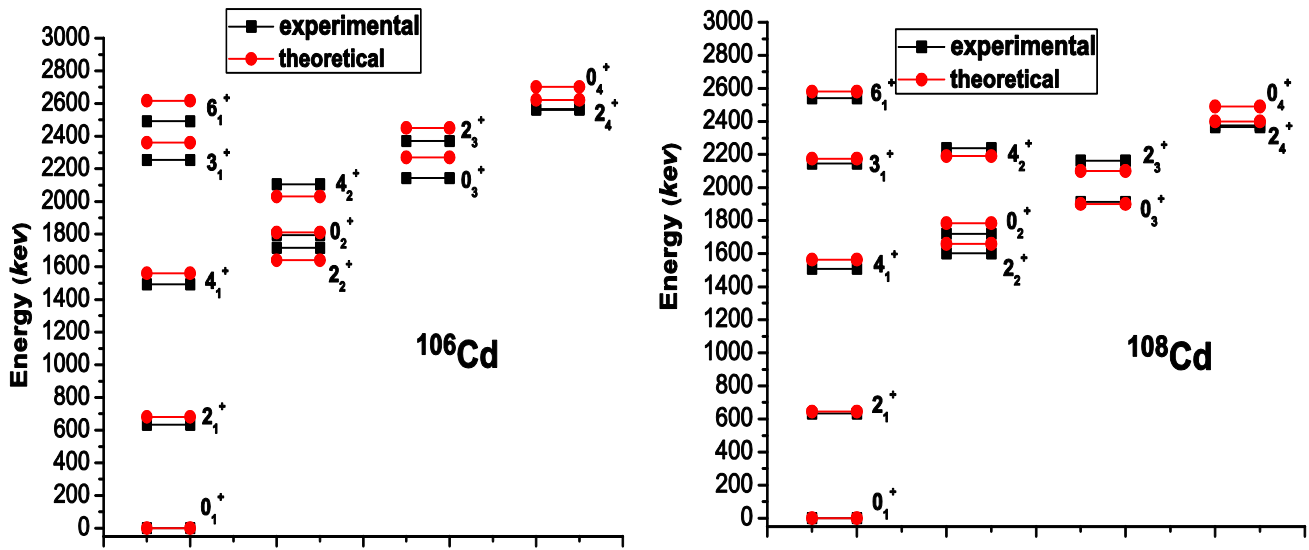

Figure2.

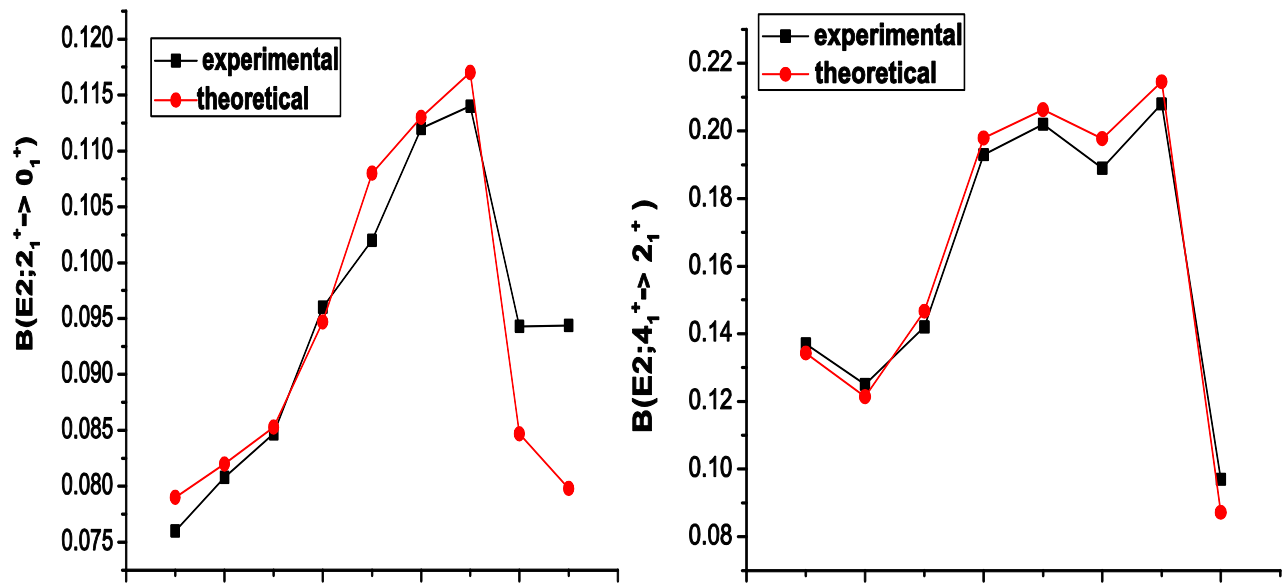

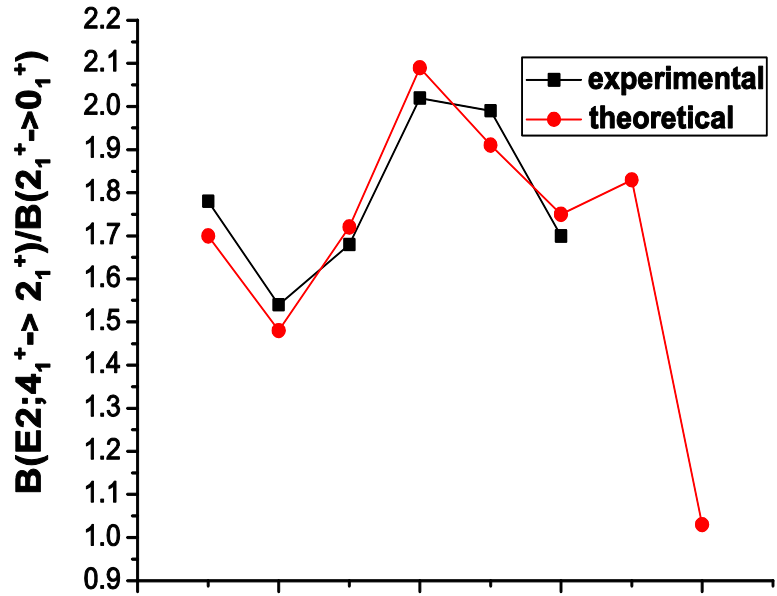

Figure3.

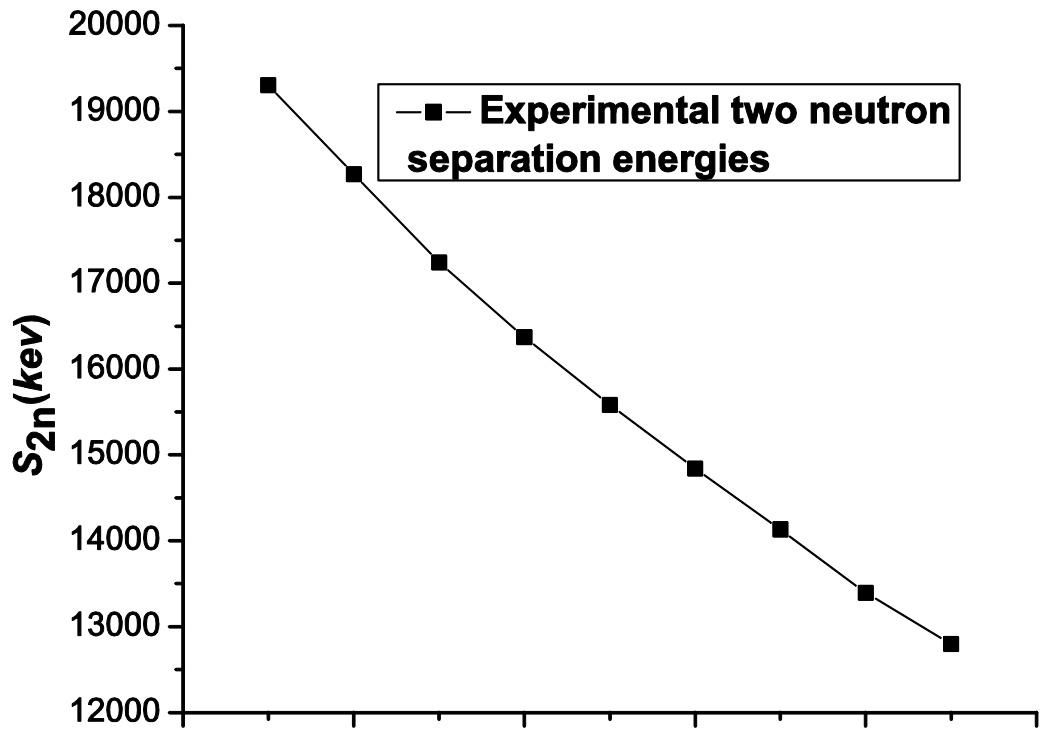